# An intelligent household greenhouse system design based on Internet of Things

**Zhonghua Han**[1,2], **Zhenbo Wu**[1], **Shuo Lin**[1] **and Fangjun Luan**[1]

[1] Faculty of Information and Control Engineering, Shenyang Jianzhu University, Shenyang 110168, China

[2] Department of Digital Factory, Shenyang Institute of Automation, the Chinese Academy of Sciences (CAS), Shenyang 110016, China

E-mail:shuaizhenbo@163.com

**Abstract.** In order to combine indoor greenery conservation with Internet of Things (IOT) Technologies, this paper designs an intelligent household greenhouse project with the features of comprehensive sensing, reliable transmission and intelligent processing. Through the analysis of functional requirements of the intelligent household greenhouse system, an intelligent household greenhouse system is designed with the functions of greenhouse environmental data detection, greenhouse environmental control regulation, data remote transmission and human-computer interaction. Its sensor layer collects environmental data in real time based on the ZigBee wireless sensor network. The network layer STM32 intelligent gateway coordinates with network server, so as to exchange data from sensor layer to application layer, and solve the problems of non-blocking of data sending and receiving as well as concurrent requests of multiple mobile terminals. The application layer is designed into two types. One is a desktop management system as a data storage and analysis center, and the other is a mobile terminal APP. At the same time, we design a communication protocol that is applicable to the interaction of the three-layer structure of the Internet of Things, with the characteristics of simplicity, stability, readability, and scalability. It can avoid the mutual influence of multi-level data exchange and ensure the correctness of data circulation. In the design, the system sensor layer ensures stable transmission of various data and instructions, and the network layer has a high degree of concurrency and real time. And various measurement and control data of the sensor layer can interact with the data of mobile-terminal equipment of the application layer. The desktop management system and mobile terminal APP can monitor greenhouse data in real time and control various actuators in the greenhouse.

## 1. Introduction

At present, intelligent greenhouses are mainly used in agricultural greenhouses. As for the proposal of household greenhouse as well as design and implementation of intelligent household greenhouse system, we can draw on research results of agricultural intelligent greenhouse, combine with the spatial limitations of household greenhouse, and simultaneously meet the requirements of nurturing various plants, easy dismounting, strong practicality, and high cost performance. In this way, we can transform intelligent greenhouse from agricultural field to home life, designing an intelligent greenhouse with reasonable cost and complete functions.

Referring to the research on structure system of the IOT at home and abroad, the intelligent household system based on the ITO adopts a layered architecture, which mainly includes the sensor layer, network layer and application layer[1-3]. Currently, most sensor layers of the IOT employ wireless transmission technology. Referring to Zhang Meng's Greenhouse Remote Monitoring System Design Scheme Based on Zigbee[4] in 2012, and Liang Ruihua's proposal to replace traditional wired networks with ZigBee technology in 2016, both of them argue that Zigbee technology have the features of low-energy consumption, reliable communication, low cost and data security, which is a feasible solution for sensor layer of intelligent household greenhouse [5].

At present, the gateway design and development for intelligent household based on the IOT focus on designing an embedded network server as the only device in the network layer. Referring to Chang Yingliang's design scheme for the IOT gateway in 2014[6], and the idea that uses wireless to connect data-acquisition sensor with the Internet, proposed by H Messer and P Alpert[7], they come up with a gateway design scheme for embedded chips in the basis of ARM11 or ARM Cortex-A8 and other high-performance cores. These schemes can basically meet the IOT requirements of data circulation, analysis and processing in the IOT system. And the cost and development of such schemes are relatively high. At the same time, there are no specific solutions proposed when the system has multiple client requests and when the data blocks. Considering the gateway server based on STM32 designed by Zhang Wenjing in 2015[8], and the embedded Web server under the uC/OS-II operating system designed by Huang Xiaofen and Liu Damao in 2006[9], we find that such design solutions are low-cost and easy-development after a comprehensive analysis. However, its processor performance cannot fully meet the functional requirements in the case of big data volume analysis and multiple client requests. By analyzing the above-mentioned research on the IOT's functional design at the network level, we propose a network layer solution for the coordination and cooperation between the intelligent gateway and the server, aiming at the problem that the current network layer design cannot balance performance and cost. The network layer gateway implements the multi-task concurrency function under the real-time operating system. In the server design, we combine with multithreading, socket communication and other technical means, realizing the data flowing across the sensor layer, network layer, and application layer.

At present, the intelligent household app control system has been widely used in China. The control terminal in intelligent household system employs Android platform equipment. The Android system provides a strong technical support and good user experience, and is easy to transplant. Consequently, the mobile phones, iPad and other mobile devices can become the control terminal of the intelligent household system, which reduces the costs and improves the usability. In 2016, Zhao Wenbing's intelligent greenhouse remote monitoring system design based on Internet of Things proposes a remote monitoring strategy for mobile terminals with Android as the operating system [10]. It has the functions of user login, data query, and greenhouse control. In 2012, Yan Wei and others bring up the application of multi-threading technology in the development of Android phones, which can avoid the non-response of applications, ensure the program transform smoothly and respond quickly[11].

According to the above researches of intelligent greenhouses based on the Internet of Things, the intelligent household greenhouse has a typical IOT hierarchy. In the system, the sensor layer mainly completes the acquisition and processing of household environment and equipment status. The principal role of the network layer is to achieve data interoperability from the underlying sensor to the user's mobile client, ensuring the reliability and real-time nature of data exchange. The application layer chiefly includes types of mobile terminals, realizing real-time data acquisition and control of household devices and related devices in the intelligent greenhouse. The system design scheme covers the following: ZigBee wireless communication network is used as a short-range wireless communication method for various sensors, executors and controllers in the sensor layer; the core of the network layer is constructed through the intelligent gateway and server, and the mobile terminal APP is used in the application layer which designs the communication protocols among layers.

## 2. Requirement Analysis of the Overall System Function

The intelligent household greenhouse system is designed for the users who want to grow flowers and plants at home. In the household environment, the indoor parameters are small in range and variation with respect to the outdoors, but they are not necessarily suitable for the growing environment of flowers and plants.

Firstly, in the process of intelligent household greenhouse design, we should first consider the problem of greenhouse placing due to the limitations of the indoor space. Most of the buildings face the southeast, making it cool in the summer and better day lighting in the winter. As a result, the greenhouse is usually placed on a south-facing balcony or room. Therefore, the intelligent household greenhouse should be flexible in size, transparent, and leak-proof.

Secondly, different plants need different growth environments. In the same greenhouse, we need to nurture plants with diverse habits. It is necessary to make full use of the limited space in the greenhouse, to properly place various plants and to customize personalized planting scheme.

Thirdly, the environmental regulation ability of greenhouses is an important performance index of greenhouses. Above all, each sensor and actuator must be rationally designed and deployed at the sensor layer, so as to collect the environmental parameters in the greenhouse in real time and make corresponding adjustments according to control instructions.

Fourthly, the network layer plays an important role in data connectivity between sensor layer and application layer, and has high real-time performance for sending and receiving data. In the light of situation that network layer needs to handle numerous network connection requests, the designed server should have concurrent processing capabilities. In the entire system, the server, on the one hand, establishes a network connection with the application layer mobile client to provide services, and on the other hand, needs to set up a connection with the gateway so as to realize bidirectional communication of data from the application layer to the gateway then to the sensor layer.

Fifthly, if users want to monitor and adjust the greenhouse environment, it is necessary to design a desktop management system with data storage and analysis functions and a mobile terminal APP that can be easily carried and operated. The user can manually and automatically control the greenhouse through the current greenhouse environmental data and the operating status of the actuator displayed on the desktop management system and the mobile terminal APP.

## 3. Overview of System Structure

The intelligent household greenhouse is designed as a self-contained greenhouse. As an absolute construction with flexible and variable size, it is attached to the building's orientation and is typically placed on a south facing balcony or room to maximize the lighting. The intelligent household greenhouse is an upper and lower two-story structure. The main body adopts a transparent acrylic removable panel, which has good light permeability, rich colors, long service life, outstanding rigidity and strength, and excellent chemical resistance.

The intelligent household greenhouse system is divided into sensor layer, network layer and application layer. The system transmits the greenhouse environmental parameters to servers and mobile terminals through ZigBee, STM32 intelligent gateways and routers. At the same time, the control instructions of the mobile terminals should be sent to the devices in the greenhouse through the server.

### 3.1. System Sensor Layer

The system sensor layer is composed of ZigBee wireless sensor control network, which realizes the collection and processing of greenhouse environmental data.

### 3.2. System Network Layer

The network layer in intelligent greenhouse system consists of STM32 intelligent gateway and server. The intelligent gateway can achieve real-time and quick interaction among data in each layer, ensuring the data circulation between the sensor layer and the application layer. The intelligent gateways have the ability to analyze and process data. They can analyze alarm data in abnormal environment detected by the sensor layer, control the actuators, handle environmental anomalies, and eliminate alarms.

*3.3. System Application Layer*

The application layer refers to the desktop management system and mobile terminal APP. The user can obtain the greenhouse environment data and actuator's working status through the desktop management system and the mobile terminal APP, and controls the actuator.

The mobile phone APP acts as a mobile control terminal. Through the APP, the user regulates greenhouse environment by adjusting the setting value of greenhouse environment (automatic mode) and actuator's gear (manual mode).

As the greenhouse environmental monitoring center and data processing center, the desktop management system can store communication-related information, such as user's information, IP addresses and port numbers, environmental data, actuators' working status, data instructions, and control instructions. The communication data between sensor layer and application layer of mobile phone APP is stored in the database of desktop management system. After being processed by the desktop management system, the data is forwarded to the gateway and the mobile phone APP.

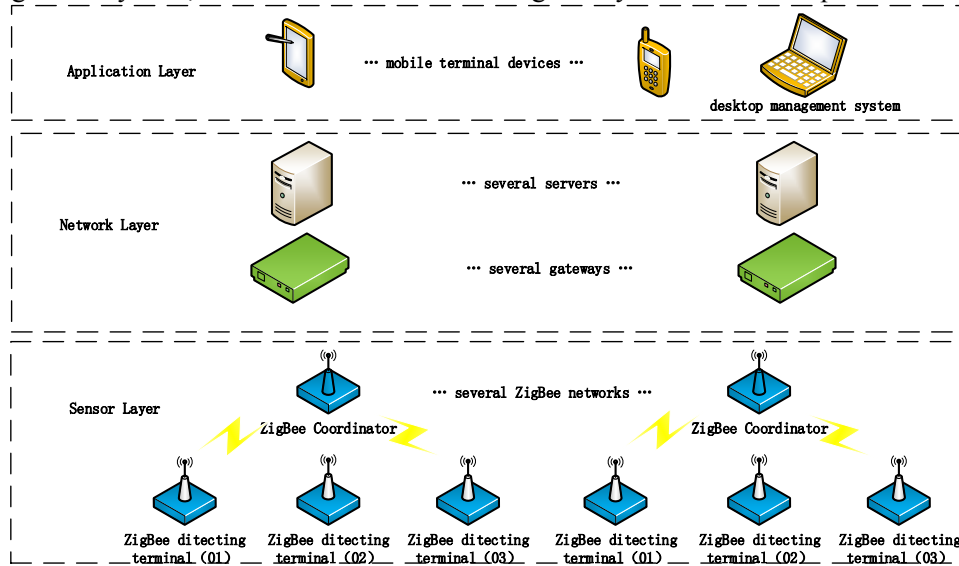

**Figure 1.** Intelligent Household Greenhouse System Structure

## 4. Overview of System Structure

*4.1. ZigBee Network Design*

The sensor layer uses ZigBee wireless technology to achieve short-range communication. The ZigBee features low-energy consumption, low cost, short delay, large network capacity, and reliable security. The ZigBee coordinator is directly connected to the ZigBee terminal nodes that are equipped with diverse sensors and actuators to establish the entire ZigBee network. In this ZigBee wireless sensor control system, the ZigBee network is designed to consist of nine ZigBee modules. One of the ZigBee modules serves as the coordinator of the ZigBee network, and the remaining eight pieces act as the terminals of the ZigBee network, six of which are detecting terminal equipped with sensors and two of which are executive terminals equipped with executors. The ZigBee communication module design includes ZigBee coordinator module, ZigBee detecting terminal module, and ZigBee executive terminal module. The core adopts CC2530 chip, and carries 8051CPU and 2.4GHZ RF transceiver. The CC2530 has different operating modes and meets the requirements for low-energy consumption.

*4.1.1. ZigBee Coordinator Design* .The software design of the ZigBee wireless sensor control system is developed by using Z-Stack protocol stack. The ZigBee coordinator is responsible for establishing the ZigBee network and interacting with the STM32 intelligent gateway in network layer. When the ZigBee coordinator receives a query sent from the serial port of the STM32 intelligent gateway in network layer, it will package the queried sensor data into a communication protocol and return it to the STM32 intelligent gateway in network layer through the serial port; when the ZigBee coordinator receives control instructions from the serial port of STM32 intelligent gateway in network layer, it will send the control instructions to the ZigBee executive terminal, and receive the executors' working status returned from the control terminal. What's more, the executor will be packaged into a communication protocol and return to STM32 intelligent gateway in network layer through the serial port.

The ZigBee coordinator module includes RS232 serial communication module, power module, reset module, Debug module, I/O and RF antennas. The CC2530 chip integrates CPU, memory, and RF transceiver, which simplifies the circuit design. And the CC2530 chip contains decoupling circuit, in order to remove the noise on the chip's power pin.

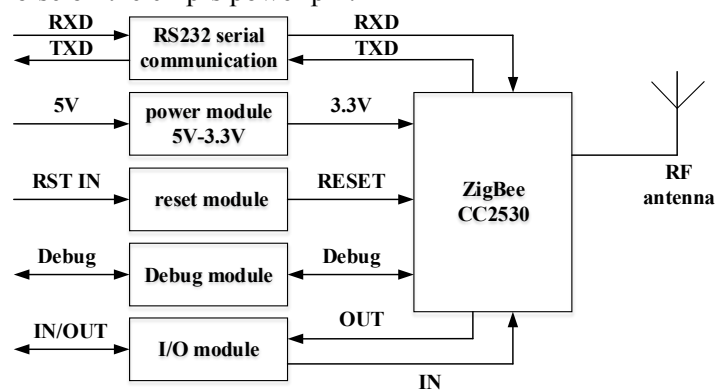

**Figure 2.** Module Structure of ZigBee Coordinator

*4.1.2. ZigBee Detecting Terminal Design* .The environmental factors that affect plant growth and can be regulated in the greenhouse are temperature, humidity, light, gas environment and soil environment. In the household greenhouse environment, the temperature ranges from -10°C to 40°C, the humidity ranging from 10% to 100%, and the light intensity ranging from 0lux to 30000lux. In order to obtain more accurate environmental parameters, we must choose sensors with right size, high accuracy and suitable range. The ZigBee detecting terminal module includes sensor input module, power module, reset module, Debug module, I/O and RF antennas.

A temperature and humidity sensor DHT11 is used to detect the temperature and humidity of the greenhouse. The DHT11 sensor is a temperature and humidity conforming sensor that contains a calibrated digital signal output, the advantages of fast response, strong anti-interference, low-energy consumption, and high cost performance. The power input VCC of the DHT11 temperature and humidity digital sensor is powered by 3.3V DC, and the reset circuit is reset by a low level. Using an external power supply, we add a 4.7kΩ pull-up resistor to the data bus of the DTH11, for the purpose of increasing circuit stability by preventing interference.

The environmental light intensity sensor of BH1750FV digital output is used to detect light intensity of the greenhouse. The BH1750FV make communication based on the I2C bus communication protocol. During data transfer, SCL is high level, while the data on the SDA line must remain stable. Only when SCL is low level are data on the SDA line allowed to change.

The greenhouse soil moisture is detected by FC-28 soil moisture sensor, which employs LM393 as a comparator to export soil moisture analogue quantity signal output and soil moisture threshold digital quantity output. The soil moisture threshold is adjusted by a potentiometer, being able to define the current soil moisture as wet and dry state. When lower than the setting value, the soil moisture is in dry output high level. When higher than the setting value, the soil moisture is in wet output low level. The STM32 directly detects the digital quantity signal of soil moisture, setting low setting of soil moisture as the alarm signal. In this ZigBee network, six detecting terminals are set, whose functions are the same, all equipped with DHT11 temperature and humidity sensor, BH1750FVI light intensity sensor, and FC-28 soil moisture sensor. These six ZigBee detecting terminals have different number which are 01~06,corresponding to six different detecting positions. And the number corresponds to the address in the communication protocol. The final data of temperature, humidity, light intensity, and soil moisture are obtained by averaging the sensor data from six different locations. After the ZigBee coordinator establishes the network, the ZigBee detecting terminal will access the network and drive the installed sensors to work. The ZigBee detecting terminal collects the environmental data in the greenhouse in real time, integrating and packaging the data and location number, and sending them to the ZigBee Coordinator.

*4.1.3. ZigBee Executive Terminal Design.* The environmental factors that affect plant growth and can be regulated in the greenhouse are temperature, humidity, light, gas environment and soil environment. In the household greenhouse environment, the temperature ranges from -10°C to 40°C, the humidity ranging from 10% to 100%, and the light intensity ranging from 0lux to 30000lux. In order to obtain more accurate environmental parameters, we must choose sensors with right size, high accuracy and suitable range. The ZigBee detecting terminal module includes sensor input module, power module, reset module, Debug module, I/O and RF antennas.

The irrigation equipment uses 12V vertical submersible pumps with 4.5W rated power. And there are 6 outlets circularly placing on the surface of the soil layer. Each floor of the greenhouse is a drawer sink, providing water for submersible pumps and humidifiers. At the same time, the excess water after irrigation will flow back to the sink through the holes at the bottom of the greenhouse, which not only solves the problem of water leakage, but also reduces the waste of water resources. The temperature of household greenhouse will not fluctuate. For one thing, the sun can provide about 25% heating. For another, according to different plants requirements for different temperatures, we will use 12V DC ceramic heating plate to warm the greenhouse, posted in the greenhouse's front and rear bottom sides. Ventilation can promote gas exchange in the greenhouse, lower the temperature and supplement co2. The heat dissipation and ventilation equipment employs a 12V, 3.9W DC fan with 8cm diameter. The heat dissipation fan is divided into an inlet end and an exhaust end. Because the hot air generally goes upward and the cold air goes down, the fan is installed in a low-inlet and high-exhaust way. Unsuitable humidity will make plants get diseases. As a result, we use 12V, 2.5W DC fan with 7cm diameter for ventilation and dehumidification, employing 24V atomization humidifier for humidification. In northern China, the duration of sunshine is short. When the light intensity is insufficient, plants will grow weak and even die. Therefore, supplementing light is one of the indispensable tasks in the greenhouse. The 5V blue-violet LED featuring with a small amount of heat, is installed on the top of the greenhouse, and the executive equipment is controlled by electric relay.

Number 07 and number 08 these two executive terminals are set in this ZigBee network. Each executive terminal is equipped with drip irrigation pump, heat dissipation fan, dehumidification fan, supplementing light LED, heating lamp, and humidifier, which can expand and increase executors. After the ZigBee coordinator creates the network, ZigBee executive terminal accesses to the ZigBee network, waiting for the broadcast from the ZigBee Coordinator. When the ZigBee coordinator receives a control command from the serial port, the ZigBee coordinator will broadcast the control instruction to the ZigBee network. After the executive terminal receives the control instruction, it will first determine whether the instruction is valid, and controls the executor under the control instruction if the determination is correct. Meanwhile, the current working state of the executor is packaged into a communication protocol and sent to the ZigBee coordinator. Then, the ZigBee coordinator sends the communication protocol to the STM32 gateway through the serial port.

*4.2. STM32 Intelligent Gateway Design*

The STM32 intelligent gateway is the core device of the intelligent household greenhouse network layer. In this intelligent household greenhouse system, the intelligent gateway is designed to have a network client that connects the ZigBee network coordinator in sensor layer and network layer server. That is, the main work of gateway is interacting data through the serial port and ZigBee network coordinator in sensor layer, which is under author's design "instruction protocol and data protocol of sensor layer, as well as data protocol of network layer sensor and device status", and analyzing different types of sensor layer data.

In the design process, full consideration should be given to the real-time requirements of the intelligent gateway in processing the data uploaded by the sensor layer and the instructions issued by the application layer. The intelligent gateway of the network layer serves as the core part of connecting sensor layer and network layer, which requires high real-time performance for sending and receiving data. The abnormal data alarm of the sensor layer is first delivered to the intelligent gateway. The identification and processing of the alarm require the characteristics of non-blocking and high realtime. At the same time, there should be no data interference with the process of sending and receiving data by the server, and there should be no process blocking[12].

In the process of designing the gateway, we must follow the idea of modular programming, respectively establishing two independent tasks of data receiving and sending of the network communication module, so as to avoid interfering with the normal work caused by blocking in the process of sending and receiving data, and ensure the real-time data circulation. We also establish an independent task to deal with data alarm in sensor layer, which is not affected by network connections and concurrently processed with other processing tasks. In this way, the alarm condition can be handled in real time. In addition, we employ interrupt mode execution to communicate with the ZigBee serial in sensor layer.

The μC/OS-II embedded operating system is transplanted in the gateway. Its function of supporting multi-task concurrency is mainly implemented by the system-supplied OSTaskCreate() API function. After the gateway completes hardware resource initialization and system initialization, a new task can be created by calling OSTaskCreate().

After the initialization of hardware, protocol stack and operating system, we will establish multi-process for concurrent processing. The ZigBee coordinator and the STM32 intelligent gateway communicate through the serial port whose data sending and receiving are performed separately. The serial port receiving data thread is an evaluation operation to the data global variable. The STM32 intelligent gateway receives the environmental parameters sent by the ZigBee coordinator and the working state of the executor, and then assigns the received data to the global variable. The serial port sending data thread reads the global variable of the control instruction, and relays the control instruction sent from the server to the ZigBee coordinator. First, we have to judge whether the flag bit is 1. If the flag bit is 1, we should read the global variable of the control instruction. Then, we should package the control instruction into a communication protocol that communicates with the ZigBee Coordinator, send it to the ZigBee Coordinator to control the executor.

The STM32 intelligent gateway communicates with the server through the network whose data sending and receiving is performed separately. The network data receiving thread is an evaluation operation to the global variable of the control instruction, receives the control instruction sent by the server to the STM32 intelligent gateway, assigns the control instruction to the global variable of the control instruction after analysis, and then sets the flag bit to 1. The network data sending thread is a reading operation to data global variable, reads data global variable, and packages the data into a communication protocol that communicates with the server, sending it to the server at regular intervals.

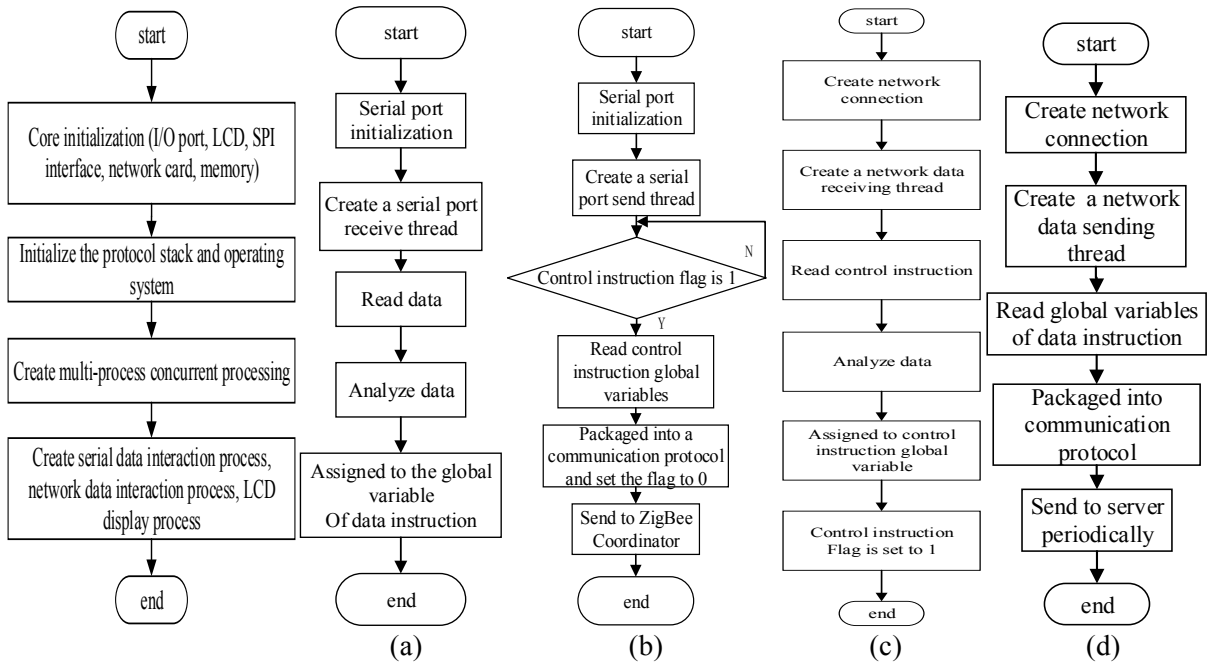

**Figure 3.** STM32 Intelligent Gateway Software Process

*4.3. Software Design of Desktop Management System*

The application layer is divided into desktop management system and mobile terminal APP. The desktop management system software of the intelligent household greenhouse designed in this paper includes user login management, greenhouse environmental monitoring, historical data statistical analysis and communication services.

The data of temperature, humidity, light intensity and soil moisture in the greenhouse obtained by sensor layer, as well as active status of actuator are periodically sent to the desktop management system for storage and updating to the interface. The manual-control instructions and automatic-control programs sent by mobile terminal APP are all stored in the database of desktop management system. Through the analysis and comparison of data, the actuators are manually and automatically controlled. The large amount of historical data in the desktop management system and the curve diagram formed from these data can facilitate the users to analyze the current control scheme and form a better control strategy.

The front-end interface design of the desktop management system uses the WPF user interface framework, which contains both program code and XAML. Firstly, we employ XAML to define the initial interface of the application, and then write the corresponding implementation code of the function. After that, we will embed the logic code directly into each XAML file[13].The database development adopts the SQL Server 2008 database which allows to use the data in custom applications developed by Microsoft .NET and Visual Studio [14]. The desktop management system database of the intelligent household greenhouse has 13 tables used to record communication-related information, such as user's information, IP addresses and port numbers, environmental data, actuators' working status, data instructions, and control instructions.

*4.3.1. Design of Desktop Management System's Function.* The software design of desktop management system uses the WPF user interface framework, containing both program code and XAML. Firstly, we employ XAML to define the initial interface of the application, and then write the corresponding function implementation code. After that, we will embed the logic code directly into each XAML file.

The user login management is applied to authenticate and log in user's account and password. When user enters the account and password, they will be compared with the information stored in the database, so as to achieve secure login.

The environmental monitoring is employed to display the real-time environmental parameters in the greenhouse and the working state of the actuators, and regulates the actuators. This module has the user management, mobile phone management, gateway management, and network settings.

The communication service program establishes a network connection with intelligent gateway client in network layer, and also needs network connection with mobile client in application layer to achieve data interaction.

The historical data analysis draws the curve of historical data by reading the data of temperature, humidity, and lighting in the database. Historical data is a true and comprehensive record of the past in the greenhouse, as well as the data source for analyzing the correlation between parameters, which is of great importance for users to maintain and manage the plants [15]. Correlation between parameters is analyzed on historical data. From the analysis results, we know that when one parameter changes, the other parameters will be influenced. Through analyzing plenty of historical data, we conclude the experience and countermeasures for plant management. And combined with growth pattern of the plant, a well-focused knowledge base is established to provide the basis for user decisions and device control.

As to the automatic-control mode of the intelligent household greenhouse system, the fuzzy neural network is used for intelligent control. In the greenhouse environment, the variables that have a greater impact on plants are temperature and humidity. Therefore, in the design of fuzzy neural network controller in this paper, the inputs are temperature error, temperature error variation, humidity error, and humidity error variation, while the outputs are temperature controlled variable and humidity controlled variable. For the temperature controlled variable, the positive value is the heating control and the negative value is the cooling control. For the humidity controlled variable, the positive value is the humidification control and the negative value is the dehumidification control.

For temperature error $e_T$, its actual range is set to [-5, 5] in this paper, and its fuzzy set domain is set to {-3, -2, -1, 0, 1, 2, 3}. And the quantitative factor of temperature error is $k_{et}=3/5$. Based on the quantitative factor, the temperature error can be mapped from the basic domain to the fuzzy set domain. At the same time, the fuzzy quantity of temperature error is designed to 7 levels, and the corresponding fuzzy set is Et = {NB, NM, NS, ZO, PS, PM, PB}.

For the temperature error variation $\Delta e_T$, its actual range is defined as [-2, 2] in this paper, and its fuzzy set domain is set to {-2, -1, 0, 1, 2}. And the quantitative factor of temperature error variation s $k_{ect}=2/2$. Based on the quantitative factor, temperature error variation can be mapped from the basic domain to the fuzzy set domain. Meanwhile, the fuzzy quantity of temperature error variation is designed to 5 levels, and the corresponding fuzzy set is ECt={NM,NS,ZO,PS,PM}.

For the humidity error $e_H$, its actual range is defined as the basic domain of the humidity error, setting to [-20%, 20%] in this paper. Its fuzzy set domain is set to {-3, -2, -1, 0, 1, 2, 3} and its quantitative factor is $k_{et}=3/20\%$. Based on the quantitative factor, the humidity errorcan be mapped from the basic domain to the fuzzy set domain. At the same time, the fuzzy quantity of the humidity error is designed to 7 levels, and the corresponding fuzzy set is Eh={NB, NM, NS, ZO, PS, PM, PB}.

For the humidity error variation $\Delta e_H$, its actual range is defined as [-5%, 5%] in this paper, and its fuzzy set domain is set to {-2, -1, 0, 1, 2}. And the quantitative factor of humidity error variation is $k_{ech}=2/5\%$. Based on the quantitative factor, humidity error variation can be mapped from the basic domain to the fuzzy set domain. At the same time, the fuzzy quantity of the humidity error variation is designed to 5 levels, and the corresponding fuzzy set is ECh= {NM, NS, ZO, PS, PM}.

In the fuzzy neural network controller of the intelligent household greenhouse system, the outputs are temperature-control adjustment and humidity-control adjustment. The temperature-control adjustments are divided into heating adjustment and cooling adjustment, and humidity-control adjustments are divided into humidification adjustment and dehumidification adjustment. The heating adjustment is a heating plate with 5-gear mode, and the cooling adjustment is a cooling fan with 5-gear mode. In addition, the humidification adjustment is a humidifier with 1-gear mode, and the dehumidification adjustment is a dehumidifying fan with 5-gear mode.

For the temperature-control adjustment, its actual range is defined as [-5, 5] in this paper, and its fuzzy set domain is set to {-3, -2, -1, 0, 1, 2, 3}. And the quantitative factor of temperature-control adjustment is $k_{tcn}=3/5$. Based on the quantitative factor, temperature-control adjustment can be mapped from the basic domain to the fuzzy set domain. At the same time, the fuzzy quantity of temperature-control adjustment is designed to 7 levels, and the corresponding fuzzy set is Tcon= {NB, NM, NS, ZO, PS, PM, PB}.

For the humidity-control adjustment, its actual range is defined as [-5, 1] in this paper, and its fuzzy set domain is set to {-3, -2, -1, 0, 1, }. And the quantitative factor of humidity-control adjustment is $k_{hcn}=3/5$. Based on the quantitative factor, humidity-control adjustment can be mapped from the basic domain to the fuzzy set domain. Meanwhile, the fuzzy quantity of humidity-control adjustment is designed to 5 levels, and the corresponding fuzzy set is Hcon= {NB, NM, NS, ZO, PS}.

The control strategy for temperature and humidity in the greenhouse is analyzed and summarized. And we design 35 control rules of the temperature adjustment and humidity adjustment respectively, as shown in Table 1 and Table 2.

**Table 1. Fuzzy rules of temperature adjustment**

| ECt | Et | | | | | | |
|---|---|---|---|---|---|---|---|
| | NB | NM | NS | ZO | PS | PM | PB |
| NM | NB | NB | NM | NS | PS | PS | PM |
| NS | NB | NM | NS | NS | PS | PM | PM |
| ZO | NM | NS | NS | ZO | PS | PM | PM |
| PS | NS | NS | NS | PS | PM | PM | PB |
| PM | NS | NS | PS | PS | PM | PB | PB |

**Table 2. Fuzzy rules of humidity adjustment**

| ECh | Eh | | | | | | |
|---|---|---|---|---|---|---|---|
| | NB | NM | NS | ZO | PS | PM | PB |
| NM | NB | NB | NM | NS | PS | PS | PS |
| NS | NB | NM | NS | NS | PS | PS | PS |
| ZO | NM | NS | NS | ZO | PS | PS | PS |
| PS | NS | NS | NS | ZO | PS | PS | PS |
| PM | NS | NS | ZO | PS | PS | PS | PS |

*4.3.2. Communication Service Program on Desktop Management System.* When the communication service program sends and receives data with gateway client and mobile client, it is required to implement real-time communication, and to avoid the mutual influence and interference between gateway client data and mobile client data. In order to meet this demand, the author puts forward using socket communication and multi-thread technology to design the server side with two sockets. One socket is used to establish a connection with the gateway client, handling the connection request of the gateway client. The procedure of desktop management system and gateway communication is shown in Figures 4(a) and (b). The other is applied to process the connection request of the mobile client. These two sockets bind the same IP address with different port number, responsible for data analysis and forwarding between each other. Meanwhile, the data, actuator's working state, control instructions, and communication data detected by the sensor will be stored. The procedure of desktop management system and mobile phone APP communication is shown in Figures 4(c) and (d). In the experimental environment, the port number connected to the gateway is 8080, and the port number connected to the mobile client is 8088. When the server communicates with these two clients, the data-receiving thread and data-sending thread are established. These two threads perform independently, so as to prevent blockage when the network-communication data are interacting.

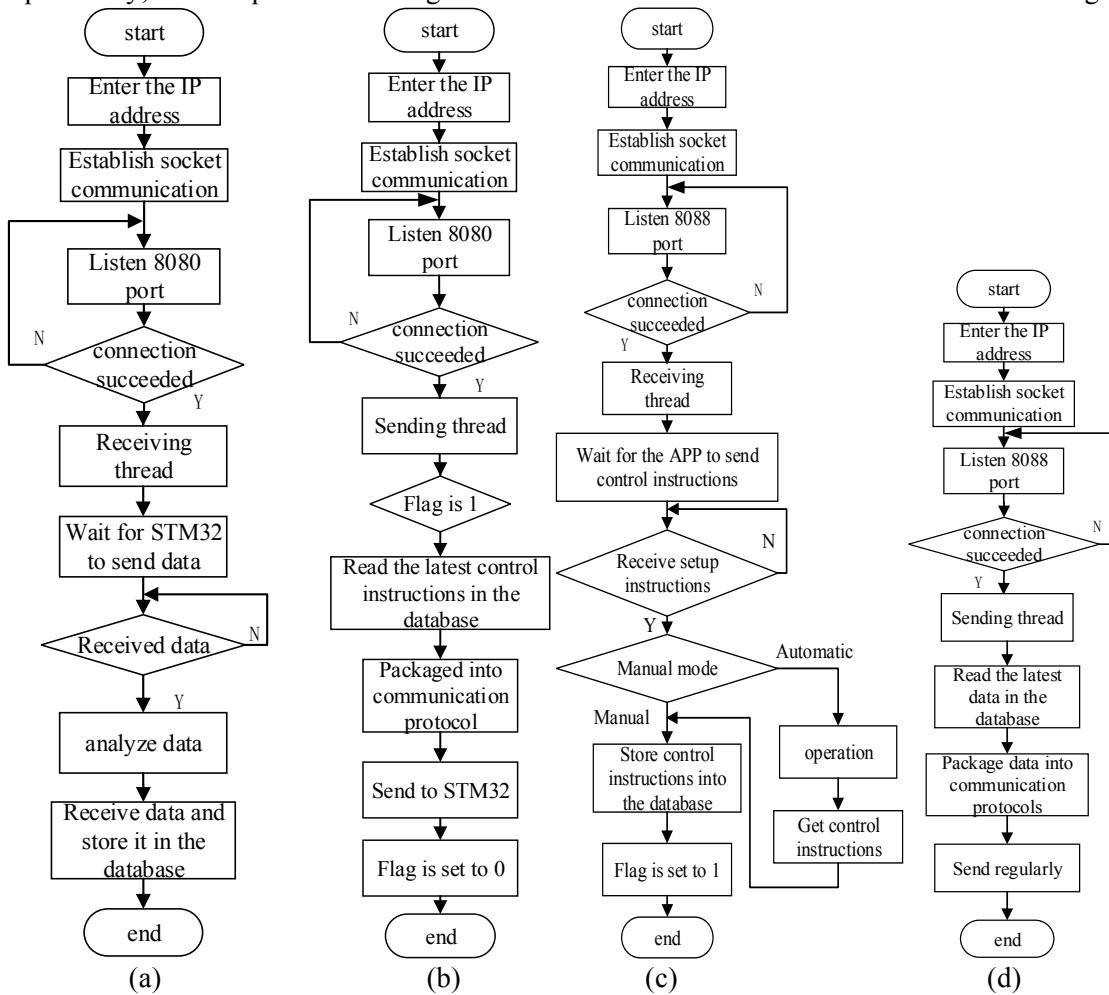

**Figure 4.** Flow Chart of Communication Service Software Program of Desktop Management System

4.4. *APP Software Design.* The App software design of system application layer is based on the Android Studio development platform. It is mainly designed by various component API interfaces provided by Android. The APP interface of intelligent household greenhouse system includes three parts: network connection interface, monitoring interface and adjustment interface.

The APP communicates with the server through the TCP-based Socket. The server first initializes the Socket, then binds to the port and monitors the port, waiting for client's connection by calling Accept(). The client creates a Socket object based on the server's IP address and port number, connecting to the server. The APP obtains the Socket output stream through the getOutputStream method provided by the Socket class, and the Socket input stream through getInputStream method, so as to interact data with the server. At the same time, in order to avoid the ANR (Application Not Responding) problem, we apply multi-threaded programming, define the handler in the main thread, and implement the handleMessage method for this handler. Afterwards, the main thread's handler is called in the child thread, and sent message through its sendMessage method[16].

When the APP is running for the first time, the user needs to add an IP address and a port number to create a socket communication to connect with the server. The information added by the user can be saved to the mobile client for later operation. After the user clicks the monitor in the Tab bar at the bottom of the App interface, then going to the monitoring interface, the APP will first judge whether the network connection is successful. If the network connection works, the APP will display the packaged communication protocol, which is sent from server at regular time, in the APP monitoring interface. The interface's adjustment function is designed for automatic and manual mode for the user to select. Under the automatic mode, according to the setting value input by the user, the APP follows an automatic control mode instruction to the server. Then the server automatically controls the actuator to perform an action through data comparison. In the manual mode, the user manually controls the executor by sliding the progress bar. Following the manual control instruction protocol from application layer, the APP sends the manual mode control instruction to the server, so as to control executor's action.

4.5. *Communication Protocol Design* . The author designs a complete set of communication protocol for the following reason: to realize data circulate from sensor layer to application layer; to avoid the mutual influence of data exchange among multiple layers; to prevent wrong data in the complex environment from interfering the system; to ensure the accuracy of data. The role of each domain in the communication protocol should be clear. The length and location of the data domain are fixed and clear. The length of the protocol is appropriate, with complete information. And the type of protocol package should be reduced, so as to improve the readability. In addition, it also includes a check mechanism so that the receiver can determine the validity and integrity of the protocol package. When adding functionality or implementing equipment in the future, the protocol will only increase in quantity without changing the structure, and still be able to perform its job. The communication protocol is designed by using six fields: frame header, lengths, addresses, types, instruction (data), and end bit.

The frame header is located in the lead byte of the protocol, used to determine the legitimacy of the data. First of all, the frame header is judged. If not comply with the protocol, both interactive sides will directly determine it as illegal data, not making process and jumping out to prepare for the next data interaction. In this communication protocol, the frame headers are all defined as 0xA5.

The length indicates the total number of bytes in data interaction process. According to the protocol, the interactive side should determine whether the received data length is consistent with that of protocol. If the data does not meet the protocol specified length, it will be judged as illegal data and not be processed.

The address represents the corresponding numbers of the detecting terminal and the executive terminal in the ZigBee network. All executors are on the same ZigBee terminal. If you want to increase or decrease executors, you only need to add or subtract the addresses.

The types respectively indicate which node of the sensor layer and the corresponding operation of the sensor layer. Different executors correspond to different properties such as temperature, humidity and lighting. Different types of data represent different properties.

The instruction (or data) indicates the commands issued to the nodes of the sensor layer. As for the executor, the instruction is the gear, and the data is the working status. For the sensor, the instruction is the query data, and the data is the detected real-time data.

The end bit includes high and low bits, namely 0x0d and 0x0a respectively, which is used to mark the end of the read data.

This part of protocol is followed by data interaction between the ZigBee node in sensor layer and the intelligent gateway in network layer through serial connection. The instruction protocol (part) of sensor layer is shown in Table 3, and data protocol (part) of sensor layer is shown in Table 4.

**Table 3.** Instruction Protocol (Part) of Sensor Layer.

| Instruction | Frame Header | Length | Address | Type | Instruction | End bit |
|---|---|---|---|---|---|---|
| supplemental lighting LED on 1st gear | A5 | 06 | 07 | 30 | 01 | 0D |
| supplemental lighting LED on 2nd gear | A5 | 06 | 07 | 30 | 02 | 0D |
| CLED | A5 | 06 | 07 | 30 | 00 | 0D |
| Heating plate on 1st gear | A5 | 06 | 07 | 31 | 01 | 0D |
| Heating plate on 2nd gear | A5 | 06 | 07 | 31 | 02 | 0D |
| Heating off | A5 | 06 | 07 | 31 | 00 | 0D |
| Collect temperature on location1 | A5 | 06 | 01 | 20 | 10 | 0D |
| Collect temperature on location 2 | A5 | 06 | 02 | 20 | 10 | 0D |

**Table 4.** Data Protocol (Part) of Sensor Layer.

| Data | Frame Header | Length | Address | Type | Data | End bit |
|---|---|---|---|---|---|---|
| supplemental lighting LED on 1st gear | A5 | 06 | 07 | 50 | 01 | 0D |
| supplemental lighting LED on 2nd gear | A5 | 06 | 07 | 50 | 02 | 0D |
| supplemental lighting LED off | A5 | 06 | 07 | 50 | 00 | 0D |
| Heating plate on 1st gear | A5 | 06 | 07 | 51 | 01 | 0D |
| Heating plate on 2nd gear | A5 | 06 | 07 | 51 | 02 | 0D |
| Heating plate off | A5 | 06 | 07 | 51 | 00 | 0D |
| temperature on location 1 | A5 | 06 | 01 | 01 | data | 0D |
| temperature on location 2 | A5 | 06 | 02 | 01 | data | 0D |

Table 5 and Table 6 are the data protocols of network layer followed by the STM32 when it sends data to the server. The "locations 1 to 6" represent the data detected by sensors at different locations. The "gear" indicates the operating capacity of the current actuator. The switch 0 means work suspension and switch 1 represents on work. Table 7 is instruction protocol of network layer followed by the server when it sends automatic and manual mode instructions to the STM32 gateway.

**Table 5.** Data Protocol of Network Layer Sensor.

| Frame Header | Length | Temperature | | Humidity | | Light intensity | | Soil moisture | | End |
|---|---|---|---|---|---|---|---|---|---|---|
| | | Type | Location1~6 | Type | Location1~6 | Type | Location 1~6 | Type | Location 1~6 | |
| A5 | 2E | 60 | data1~6 | 61 | data1~6 | 62 | data1~6 | 63 | data1~6 | 0D |

**Table 6.** Data Protocol of Executor Status in Network Layer.

| Frame Header | Length | LED | | Heating plate | | Cooling fan | | Dehumidify fan | | Drip irrigation system | | Humidifier | | End |
|---|---|---|---|---|---|---|---|---|---|---|---|---|---|---|
| | | Type | Gear 0~3 | Type | Gear 0~5 | Type | Gear 0~5 | Type | Gear 0~5 | Type | Gear 0~1 | Type | Switch 0~1 | |
| A5 | 2E | 50 | 00~03 | 51 | 00~05 | 52 | 00~05 | 53 | 00~05 | 54 | 00~01 | 55 | 00~01 | 0D |

**Table 7.** Instruction protocol of Network layer.

| Frame Header | Length | LED | | Heating plate | | Cooling fan | | Dehumidify fan | | Drip irrigation system | | Humidifier | | End |
|---|---|---|---|---|---|---|---|---|---|---|---|---|---|---|
| | | Type | Gear 0~3 | Type | Gear 0~5 | Type | Gear 0~5 | Type | Gear 0~5 | Type | Switch 0~1 | Type | Switch 0~1 | |
| A5 | 0E | 30 | 00~03 | 31 | 00~05 | 32 | 00~05 | 33 | 00~05 | 34 | 00~01 | 35 | 00~01 | 0D |

Table 8 is the data protocol of application layer followed by the server when it sends data to the APP, where "result" represents the average of the following temperature, humidity, lighting and soil moisture. Table 9 and Table 10 is the instruction protocol of application layer followed by the mobile phone APP when it sends the automatic and manual mode control instruction protocol to the server.

**Table 8.** Data Protocol of Application Layer.

| Frame Header | Length | LED | | Heating plate | | Cooling fan | | Dehumidify fan | | Drip irrigation system | | Humidifier | | Temperature | | humidity | | Light intensity | | Soil moisture | | End |
|---|---|---|---|---|---|---|---|---|---|---|---|---|---|---|---|---|---|---|---|---|---|---|
| | | Type | Gear 0~3 | Type | Gear 0~5 | Type | Gear 0~5 | Type | Gear 0~5 | Type | Switch 01 | Type | Switch 0~1 | Type | res | Type | res | Type | res | Type | res | |
| A5 | 1A | 50 | 00~03 | 51 | 00~05 | 52 | 00~05 | 53 | 00~05 | 54 | 00~01 | 55 | 00~01 | 60 | data | 61 | data | 62 | data | 63 | data | 0D |

**Table 9.** Automatic Control Instruction Protocol of Application Layer.

| Frame Header | Length | Temperature | | Humidity | | Light intensity | | End |
|---|---|---|---|---|---|---|---|---|
| | | Type | Setting value | Type | Setting value | Type | Setting value | |
| A5 | 09 | 40 | data | 41 | data | 42 | data | 0D |

**Table 10.** Manual Control Instruction Protocol of Application Layer.

| Frame Header | Length | LED | | Heating plate | | Cooling fan | | Dehumidify fan | | Drip irrigation system | | Humidifier | | End |
|---|---|---|---|---|---|---|---|---|---|---|---|---|---|---|
| | | Type | Gear 0-3 | Type | Gear 0-5 | Type | Gear 0-5 | Type | Gear 0-5 | Type | Switch 0~1 | Type | Switch 0~1 | |
| A5 | 0E | 30 | 00~03 | 31 | 00~05 | 32 | 00~05 | 33 | 00~05 | 34 | 00~01 | 35 | 00~01 | 0D |

## 5. Conclusion

The article analyzes the users' needs of the intelligent household greenhouse system, proposing a design scheme with greenhouse environmental data detection and transmission, human-computer interaction system, greenhouse environmental control and regulation functions. In the meanwhile, we also consider the advantages of low cost, easy installation and aesthetics. The intelligent household greenhouse system can realize the real-time and accuracy of data collection through the design of the ZigBee network. The network layer where the intelligent gateway and server coordinate and cooperate has a high degree of concurrency and realtime, satisfying the functions of multi-client requests and data storage. The system can monitor greenhouse through desktop management system and mobile terminal APP, with manual and automatic control. At the same time, the author standardizes the communication protocol to avoid the impact of multi-layer data interaction, which can effectively handle the data interaction among multiple layers. Through the experiment, the system performance is reliable and stable. With the continuous development of the Internet of Things and intelligent household, the intelligent household greenhouse has a broad space for development.

**Acknowledgments**
This work was supported by Project of Liaoning Province Education Department（LJZ2017015）and Shenyang Science and Technology Planning Project ( Z18-5-015).